\documentstyle[12pt]{article}
\begin{document}
\begin{titlepage}
\pagestyle{empty}
\baselineskip=21pt
\begin{center}

{\large{\bf On classical anisotropies in models of open inflation}}

{Jaume Garriga$^{1,2}$ and V. F. Mukhanov{$^3$}}

{\it $^1$ IFAE, Departament de F\'\i sica,
Universitat Aut\`onoma de Barcelona,}
{\it 08193 Bellaterra (Barcelona), Spain.;}

{\it $^2$ Center For Theoretical Physics, Laboratory for Nuclear
Science}
{\it and Department of Physics, Massachusetts Institute of Technology}
{\it Cambridge, MA 02139;}

{\it $^3$ Institut f\"ur Theoretische Physik, ETH Z\"urich, CH-8093}
{\it Z\"urich, Switzerland}

\end{center}

\begin{abstract}

In the simplest model of open inflation there are two inflaton
fields decoupled from each other.
One of them, the tunneling field, produces a first stage of inflation
which prepares the ground
for the nucleation of a highly symmetric bubble. The other, a free
field,
drives a second period  of slow roll inflation inside the bubble.
However, the
second field also evolves during the first stage of inflation, which to
some
extent breaks the needed symmetry. We show that this generates large
supercurvature anisotropies which, together with the results of Tanaka
and Sasaki, rule out this class of simple models (unless, of course,
$\Omega_0$
is sufficiently close to one.) The problem does not arise in 
modified models where the second field does not evolve in the first stage 
of inflation. 

\end{abstract}
\end{titlepage}

Open inflation has recently received some attention as a viable model
for the
early universe \cite{open}--\cite{grli}. The first version of the open universe
model was based on the theory of one scalar field \cite{open,BGT}. However, no
realistic models of that type have been proposed so far.
 Then Linde and
Mezhlumian proposed a class of models involving two scalar fields \cite{lime}.
The simplest model   discussed in \cite{lime} describes a tunneling
field $\sigma$,
responsible for bubble
nucleation, and a free field $\phi$ of mass $m$ that undergoes slow
rollover.
The two fields are decoupled from each other, except of course
gravitationally. When the field $\sigma$ is in its false
vacuum, it dominates the energy density and the universe is in a de
Sitter
phase with constant Hubble rate $H_1$. The true vacuum of the the field
$\sigma$ has vanishing energy density, and when a bubble nucleates, the
slowly rolling field $\phi$
drives a second period of inflation in its interior, with hubble rate
$H_2=(m^2 \phi^2)^{1/2}$.

In order for this model to work, the field $\phi$ has to
evolve very slowly outside the bubble. Otherwise the surfaces of
constant
$\phi$ would not be well synchronized with the hyperboloids of constant
$\sigma$ inside the bubble, and large anisotropies could be expected.
The danger of this effect was already realised  in \cite{lime}, therefore Linde
and Mezhlumian 
also suggested the possible modifications of the simplest model where
synchronization was exact and no such problems appear. However since the
simplest model looks more natural, it would still be 
interesting to clarify whether it is compatible 
with observations for some range of the parameters or not.

Outside the bubble of the field $\sigma$, we have
\begin{equation}
 \phi({\bf x}, t)\approx \phi_s\equiv A e^{-\alpha H_1 \hat t},
\label{roll}
\end{equation}
where $\hat t$ is the
cosmological time in the flat Friedmann Robertson Walker (FRW) chart,
and $\alpha= m^2/3{H_1}^2 << 1$ is
the slow-rollover parameter.
We shall now concentrate on fluctuations which will arise because of
the $\hat t$
dependence of $\phi$, so we take $A$ to be constant over
the region of interest.

The true and false vacua for the field $\sigma$ are strongly
non-degenerate, which means we are in the thick wall regime..
For simplicity, we shall
take the solution (\ref{roll}) to be valid everywhere outside the
forward
light-cone from the nucleation event. This approximation is clearly
valid if the bubble size at the moment of nucleation is small compared
with
$H_1^{-1}$.

Inside the light-cone, the spacetime can be covered with the open chart
\begin{equation}
ds^2=-dt^2+a^2(t)[dr^2+\sinh^2 r(d\theta^2+\sin^2 \theta d\varphi^2)].
\label{open}
\end{equation}
The scale factor $a$ obeys the Friedmann equation
\begin{equation}
(1-\Omega)\dot a^2=1, \label{friedmann}
\end{equation}
where $\Omega$ is the density parameter and at early times $a(t)\approx
t$.
To propagate the solution (\ref{roll}) to the inside of the light-cone
we can
use the following trick. Let us assume that inside the
bubble, and for $t<t_*<<H_1^{-1}$,
the universe is still de Sitter, with the same Hubble parameter $H_1$.
Since at early times the universe is
dominated by curvature, this assumption will turn out to be harmless.
Introducing in Eq. (\ref{roll}) the relation between the flat
and open de Sitter coordinates
$\hat t=\ln(\cosh H_1t +\sinh H_1t \cosh r)$, we have
\begin{equation}
\phi_s = A (\cosh H_1 t+ \sinh H_1 t \cosh r)^{-\alpha}.\label{rollopen}
\end{equation}
Of course (\ref{rollopen}) satisfies the field equation to first order
in the
slow-rollover parameter $\alpha$
\begin{equation}
\ddot \phi_s + 3{\dot a\over a} \dot \phi_s +
{(\partial_r^2+2\coth r\partial r)\over a^2 }\phi_s +m^2\phi_s =
O(\alpha m^2\phi_s), \label{eomphi}
\end{equation}
with $a=H_1^{-1}\sinh(H_1 t)$. However, we should note that from the
point of
view of the open chart, (\ref{rollopen}) is not a slow-rollover
solution. The
largest terms in (\ref{eomphi}) for $t<<H_1^{-1}$ are the gradient and
friction terms, which balance each other.

The question is now to evolve the approximate solution
\begin{equation}
\phi_s\approx (1+H_1 t \cosh r)^{-\alpha},\label{approx}
\end{equation}
which is valid for $t<t_*<<H_1^{-1}$, into the second period of
inflation.
Since (\ref{approx}) is not  an eigenfunction
of the Laplacian, one might be tempted to expand it
treating $H_1 t_* \cosh r$ as a small quantity,
given that $t_*$ can be chosen as small as desired. However this would
not be
appropriate. The scale factor behaves as $a(t)\approx t$ up
to a time $t\sim H_2^{-1}$, where $H_2<<H_1$ is the Hubble rate
at the beginning of the second
period of inflation. Therefore, the comoving size of
our causal past (i.e. the region that has an influence
on our observable universe) at the time $t_*$ is
\begin{equation}
\Delta r \sim -\ln (H_2 t_*) >> 1 \label{width}
\end{equation}
and $\Delta(H_1 t_* \cosh r)\sim H_1/H_2 >>1$. Hence the second term in
the
parenthesis in Eq. (\ref{approx}) is not uniformly small compared to the
first
over a sufficiently large patch.

A better strategy is to study a region where the second term in
(\ref{approx})
dominates. Note that since (\ref{approx})
is spherically symmetric, the universe would still look isotropic around
us
if we lived in the privileged position $r=0$.
The most interesting effects will occur
if we live at some $r=r_0>>1$. Actually, since the volume grows
exponentially
with $r$, that is where we are most likely to be found. Taking
$r_0>-\ln(\alpha {t_*}^2H_1H_2)$, we have
$$
\phi_s\approx A(1+H_1 t_*\cosh r)^{-\alpha}\approx A(H_1 t_*\cosh
r)^{-\alpha}
(1+O(\alpha^2))
$$
over a region of comoving size (\ref{width}) centered around $r_0$.

It is now convenient to change the coordinates $(r,\theta,\phi)$ on the
spacelike hyperboloid to a new set $(r',\theta', \phi')$ such that the
point $r=r_0$ will be the new origin of coordinates $r'=0$.
One can show that
$$
\cosh r=\sinh r_0 \sinh r' \cos \theta' + \cosh r_0 \cosh r'.
$$
Our solution is now of the form
\begin{equation}
\phi_s\approx B(t) f^{-\alpha}(r',\theta'),\label{factor}
\end{equation}
where
\begin{equation}
f(r',\theta')\equiv (\cosh r' + \sinh r' \cos \theta').
\label{f}
\end{equation}
Note that, to first order in $\alpha$, $f^{-\alpha}$ is an
approximate eigenfunction of the Laplacian (with eigenvalue $-2\alpha$),
and therefore we can factorize the time dependence in (\ref{factor}).

It can be checked that between the times $t_*$ and $t_2\sim H_2^{-1}$
the
gravitational backreaction of the scalar field perturbations can be
neglected,
and hence all that will happen is that $B(t)$ will evolve as
$t^{-\alpha}$.
After $t_2\sim H_2^{-1}$ slow rollover will set in and gravitational
backreaction
has to be taken into account. For this purpose, one has to separate the
field
into a background part $\phi_0(t)$ and a perturbation $\delta\phi$.
Since the
gradient terms are negligible in the slow rollover regime (the
wavelength
of the perturbations we consider is much larger than the horizon),
$\delta\phi$ need not be an eigenfunction of the Laplacian, and the
decomposition into background and
perturbation has some degree of arbitrariness. We choose $\phi_0=B(t_2)$
and
$\delta\phi= B(t_2) \alpha \ln f$ at the time when the slow rollover
starts
inside the bubble. The standard procedure then yields the answer that at
the
end of inflation the gauge invariant potential is given by
\begin{equation}
\Phi\approx {3\over 5} \left.{H_2\delta\phi \over
\dot\phi_0}\right|_{t=t_2}=
{3\over 5}\left({H_2\over H_1}\right)^2\ln f(r',\theta').
\label{potential}
\end{equation}
Somewhat surprisingly, the answer does not depend on $m^2$ but only
on the ratio between the Hubble rates inside and outside the bubble
(of course, in a slow roll model, $H_2$ indirectly depends on $m^2$,
for fixed spatial curvature.)
In Fig. 1 we plot the potential $\Phi$ at the
surface of last scattering $r_{ls}=\cosh^{-1}[(2/\Omega_0)-1]$ for
different values of $\Omega$

It remains to be seen how $\Phi$ contributes to observables.
The present Hubble radius $H_0^{-1}$ corresponds to the co-moving
distance $r_0'=[a(t_0)H_0]^{-1}=(1-\Omega_0)^{1/2}$, where the subindex
zero
indicates the present time, and we have used (\ref{friedmann}).
If the present spatial curvature is small, then $r_0<<1$ and $\ln f$ can
be
expanded as
$$
\ln f\approx x-{x^2\over 2}+{x^3\over 3}+...
$$
where
\begin{equation}
x\equiv (\sinh r'\cos \theta'+\cosh r') -1 << 1.\label{x}
\end{equation}
As mentioned before, $\ln f$ is not an eigenfunction of the Laplacian,
and
is not normalizable, another illustration of how
non-normalizable supercurvature modes can arise in the open inflationary
models.

The dominant contribution for small curvature is
proportional to $x$. The term in parentheses in equation (\ref{x}) is a
supercurvature mode with eigenvalue of the Laplacian equal to +3, and
it can actually be checked that it is pure gauge. It corresponds to the
shift $(\delta t,\delta x^i)=(\Phi t, \Phi,_i \ln t)$
during the curvature dominated period.
The constant $-1$ in (\ref{x}) will also not induce any anisotropies.
Therefore, the first nontrivial contribution will stem from the $x^2$
term
in the expansion of $\ln f$. This is proportional to $r'^2
\cos^2\theta'$.
Hence, the dominant effect will in the quadrupole. Higher
multipoles, labeled by $l$, will be accompanied by $r'^l$. Since
$r_{ls}'\approx 2(1-\Omega_0)^{1/2}$, we have
\begin{equation}
\left.{\delta T \over T}
\right|^{class}_l\sim f_l(1-\Omega_0)^{l/2} \left({H_2\over
H_1}\right)^2,
\label{effect}
\end{equation}
where $f_l\sim 10^{-1}$.
We have numerically checked that the above estimate works well
by order of magnitude even for $\Omega_0$ as low as $.3$ and 
$l{\buildrel <\over\sim} 10$.
Notice that this is a classical effect, and cosmic variance cannot be
invoked to minimize it.
The integrated Sachs-Wolfe effect is shown in Fig. 2 as a function of the
polar angle $\theta'$ and for different values
of $\Omega_0$  (the monopole and dipole contributions have been
subtracted.)
For low $\Omega_0$, the anisotropy would show up as a cold spot of small
angular size.

Unless the curvature of the universe is
very small, the ratio $H_1/H_2$ would have to be large in order to avoid
conflict with observations.
For moderate curvature, we need $(H_1/H_2)
{\buildrel > \over \sim} 10^2$, whereas for $(1-\Omega_0)<<1$
one can get away with a lower ratio.

However, Tanaka and Sasaki have shown that
the quantum fluctuations of $\phi$ generated during the first stage of
inflation induce an enhancement of the temperature anisotropies in
supercurvature modes compared with the fluctuations
in subcurvature modes \cite{tasa,gali}
$$
\left.{\delta T\over T}\right|^{sup}_l\sim (1-\Omega_0)^{l/2}
\left({H_1\over H_2}\right)
\left.{\delta T\over T}\right|^{sub}.
$$
Comparing with equation (\ref{effect}) the ratio of Hubble constants
appears reversed.
If what we observe is due to quantum fluctuations generated during
inflation, then $\delta T/T|_{sub}\sim 10^{-5}$. With $(H_1/H_2)
{\buildrel > \over \sim} 10^2$, as required above for moderate
curvature, the
model would give rise to temperature anisotropies of order
$10^{-3}$, which are ruled out by observations.
Of course the model can be saved if the curvature is sufficiently small,
say $(1-\Omega_0){\buildrel < \over \sim} 10^{-2}$.

Also,as we already mentioned, there are alternative models \cite{lime,grli} 
in which the slow roll
field $\phi$ does not evolve during the first stage of inflation. These
would not be constrained by the mechanism discussed above.

\section*{acknowledgements}

We would like to thank T. Tanaka and A. Linde for useful comments.
This work is partially supported by SNF, by the U.S. Department of 
Energy (D.O.E.) under cooperative research agreement DE-FC02-94ER40818,
and by CICYT under project No. AEN95-0882. V.F.M. thanks the Tomalla
foundation for finantial support.

\section*{Figure captions}
\begin{itemize}
\item{Fig. 1} The gauge potential $\Phi$ at the
surface of last scattering $r_{ls}=\cosh^{-1}[(2/\Omega_0)-1]$ as a function 
of the polar angle $\theta'$, for
$\Omega_0$ ranging from .3 to .9. (The anisotropy has azimutal symmetry) 

\item{Fig. 2} The temperature anisotropy as a function of the polar
angle $\theta'$ , for the same values of $\Omega_0$ as in Fig. 1.  
This is found from
numerical integration of the Sachs-Wolfe effect along the line
of sight. The monopole and dipole have been subtracted.  

\end{itemize}

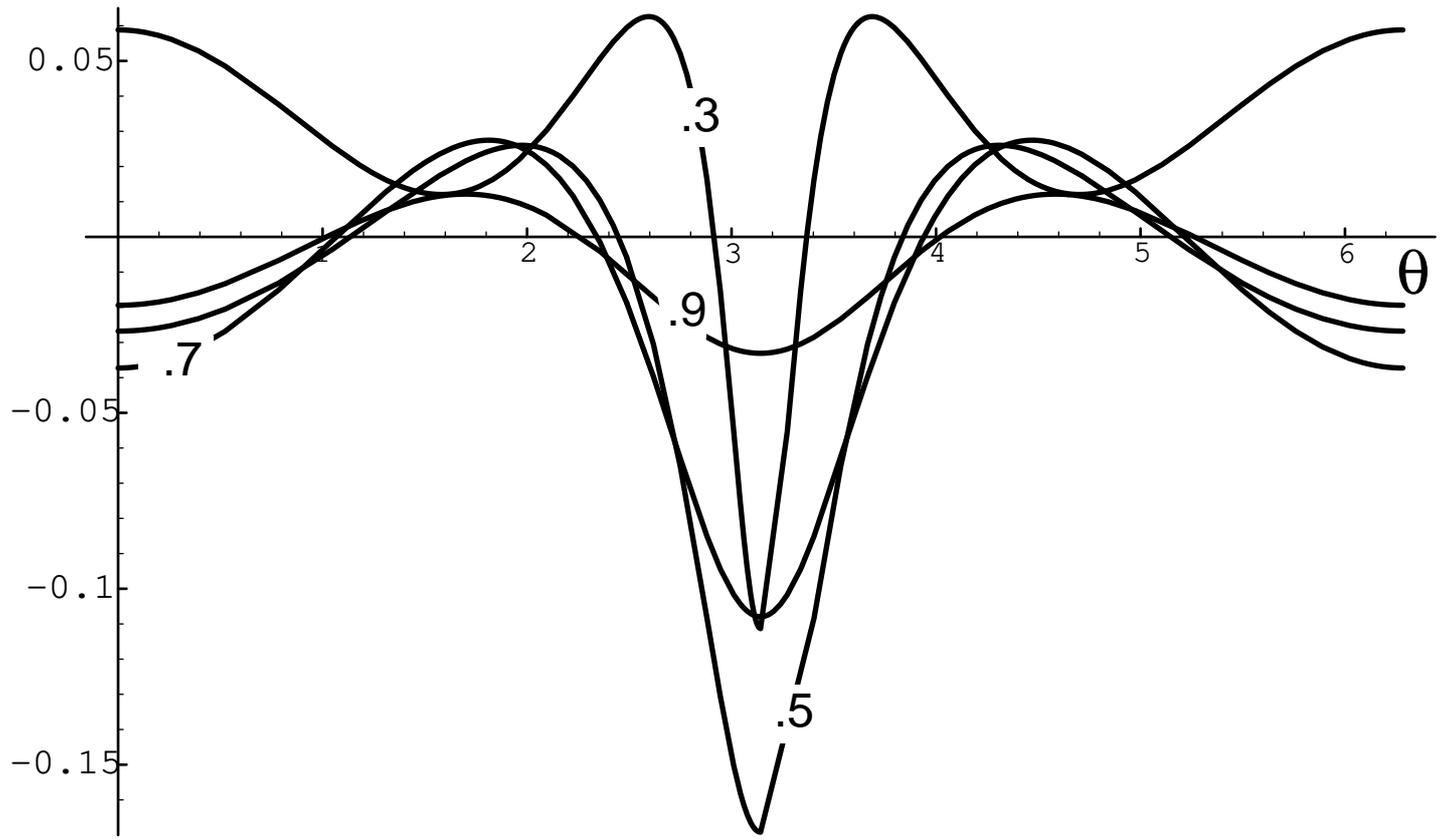
\end{document}